\documentclass[aps,prl,showpacs,twocolumn,floatfix]{revtex4}
\usepackage{latexsym}
\usepackage{amsfonts}
\usepackage{amssymb}
\usepackage{amsmath}
\usepackage{graphicx}

\begin{document}

\title{Synchronization of spin-transfer oscillators driven by stimulated microwave currents}

\author{J. Grollier}
\affiliation{Institut d'Electronique Fondamentale, UMR 8622 CNRS,
Universit\'e Paris-Sud XI, Batiment 220, 91405 Orsay, France}

\author{V. Cros}
\affiliation{Unit\'e Mixte de Physique CNRS/Thales, Route
d\'epartementale 128, 91767 Palaiseau Cedex and Universit\'e
Paris-Sud XI, 91405, Orsay, France}

\author{A. Fert}
\affiliation{Unit\'e Mixte de Physique CNRS/Thales, Route
d\'epartementale 128, 91767 Palaiseau Cedex and Universit\'e
Paris-Sud XI, 91405, Orsay, France}

\date{\today}

\begin{abstract}

We have simulated the non-linear dynamics of networks of
spin-transfer oscillators. The oscillators are magnetically
uncoupled but electrically connected in series. We use a modified
Landau-Lifschitz-Gilbert equation to describe the motion of each
oscillator in the presence of the oscillations of all the others.
We show that the oscillators of the network can be locked not only
in frequency but also in phase. The coupling is due to the
microwave components of the current induced in each oscillator by
the oscillations in all the other oscillators. Our results show
how the emitted microwave power of spin-transfer oscillators can
be considerably enhanced by current-induced synchronization in an
electrically connected network. We also discuss the possible
application of our synchronization mechanism to the interpretation
of the surprisingly narrow microwave spectrum in some experiments
on a single isolated spin-transfer oscillator.

\end{abstract}

\maketitle

The spin transfer phenomenon, predicted by J.
Slonczewski\cite{Slonc} in 1996, is now the subject of extensive
experimental\cite{Tsoi,Wegrove,Katine,JulieAPL,Myers,Ozyilmaz,Urazdhin,KiselevPRL}
and theoretical
studies\cite{Berger,Stiles,Zhang,Manschot,Fert,Barnas}. It has
first been shown that a spin polarized current injected into a
thin ferromagnetic layer can switch its magnetization. This occurs
for current densities of the order of
 10$^7$ A.cm$^{-2}$ and the switching can be extremely fast ($<$ 200 ps)\cite{Thibault}.
 More recently,
 it has been experimentally demonstrated that, under
certain conditions of applied field and current density, a spin
polarized dc current induces a steady precession of the
magnetization at GHz frequencies\cite{Kiselev,Rippard,Krivotorov}.
These steady precession effects can be obtained in
F$_{1}$/NM/F$_{2}$ standard trilayers in which a thick magnetic
layer F$_{1}$ with a fixed magnetization is used to prepare the
spin polarized current that is injected in a free thin
 magnetic layer F$_{2}$. The Giant
Magnetoresistance effect (GMR)\cite{Baibich} of the magnetic
trilayer converts the magnetic precession into microwave
electrical signals. We will refer to these non-linear oscillators
as "spin transfer oscillators" (STO).
 They
emit at frequencies which depend on field and dc current, and can
present very narrow frequency linewidths\cite{RippardPRB}. As a
consequence, they are promising candidates for applications in
telecommunications, where the need for efficient, integrated and
frequency agile oscillators is growing. The main drawback of the
spin transfer oscillator is its very weak output microwave power,
that can be optimistically estimated at - 40 dBm for a single
oscillator. A solution to overcome this difficulty is to
synchronize several oscillators, i.e. to force them to emit at a
common frequency and in phase in spite of the intrinsic dispersion
of their individual frequencies. This is essential for
applications and, on the other hand, this raises complex problems
which are new in spintronics and related to the general field of
the dynamics of non-linear systems.

Synchronization has been extensively studied since the 80's, not
only because of its many potential applications (in physics,
biology and chemistry) but also because understanding the behavior
of a large collection of non-linear dynamic systems is a
theoretical challenge \cite{StrogatzNonLin,Winfree}. In solid
state physics, a well known example of synchronization is given by
a network of Josephson junctions. An alternating potential takes
place across a single superconductor/insulator/superconductor
junction if a dc current exceeding a critical current is injected
through it. For an array of such junctions, electrically connected
in series or in parallel, each junction emits a microwave current
that adds to the injected dc current. When the resulting
interaction exceeds a critical level, it tends to synchronize the
oscillation of the junctions
\cite{Newrock,StrogatzJosephson,JosephExp}. The theoretical
prediction\cite{Duzer} is that for N oscillators, not only the
emitted power increases as N$^2$, but the frequency linewidth
decreases as N$^{-2}$. There is a definite similarity between
networks of Josephson junctions and of STO's, in spite of the
different equations ruling these two systems. Recent experiments
have shown that STO's can phase-lock (synchronize) to an external
microwave current source \cite{RippardLock}. Slavin \textit{et
al.} have analytically studied this case for weakly non-linear
spin transfer oscillators \cite{Slavin}. Even more recently, it
has been shown experimentally that two nano-contact spin transfer
oscillators can synchronize, but the origin of the coupling is
still debated \cite{Rippard2sync,Mancoff2sync}.

In this letter, we develop numerical simulations to study the
synchronization of STO's. More specifically, for spin transfer
oscillators electrically connected in series (or in parallel), we
introduce the coupling due to the microwave current induced in
each oscillator by the oscillations of all the others. We show
that, under certain conditions for the dispersion of the
frequencies, the GMR amplitude and the delay between the magnetic
precession and the current oscillation, synchronization can be
obtained with an output power increasing as N$^{2}$  for a
collection of N oscillators.

\begin{figure}[h]
   \centering
    \includegraphics[keepaspectratio=1,width=8.5 cm]{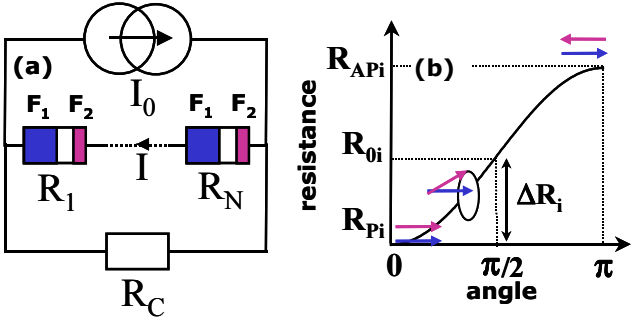}
     \caption{(a) Sketch of N oscillators connected in series and
     coupled to a load R$_C$ (throughout the paper, R$_C$ = 50
     $\Omega$).
       (b) Variation of the resistance versus the
     angle $\theta_i$ between the magnetizations of F$_{1}$ and F$_{2}$
     for an oscillator and corresponding notations.}
    \label{fig1}
\end{figure}

We first consider N oscillators of standard structure for spin
transfer F$_{1}$(fixed)/NM/F$_{2}$(free) connected in series and
coupled to a dc current generator and to a resistive load R$_{C}$,
as shown on Fig.\ref{fig1}(a). Our notation is displayed on
Fig\ref{fig1}(b). We call R$_{Pi}$ and R$_{APi}$ the resistances
of the oscillator \textit{i} in respectively its parallel and
antiparallel magnetic configurations. We define
R$_{0i}$=(R$_{APi}$+R$_{Pi}$)/2, $\Delta
R_{i}$=(R$_{APi}$-R$_{Pi}$)/2, $\beta$$_R$ = R$_{C}$ /
(R$_{C}$+$\sum_{i=1}^{N}$R$_{0i}$), and $\beta$$_{\Delta Ri }$ =
$\Delta$R$_{i}$ / (R$_{C}$+$\sum_{i=1}^{N}$R$_{0i}$). For the
dependence of the resistance R$_{i}$ of the oscillator \textit{i}
on the angle between the magnetizations of F$_{1}$ and F$_{2}$ at
time \textit{t}, $\theta_i$(t), we assume the following standard
equation :
\begin{equation}\label{resist}
R_{i} = R_{0i} - \Delta R_{i} cos[\theta_{i}(t)]
\end{equation}

The angle $\theta_i$(t) depends on the initial value of $\theta_i$
at t = 0 and on the variation of the current between 0 and $t$.

In first order of $\sum$$\beta$$_{\Delta}$$_R$$_i$ and with the
notations of Fig.\ref{fig1}(a), a straightforward calculation
leads to  :

\begin{equation}\label{hyper}
I = I_{1} + \sum_{i=1}^{N} I_{1} \: \beta_{\Delta Ri} \:
cos(\theta_i(t))
\end{equation}

with $I_{1}$ = $\beta$$_R I_0$. Similar expressions can be found
for oscillators connected in parallel, with different expressions
for J and $\beta_{\Delta Ri}$.

In order to study the behavior of N electrically coupled
oscillators, we have performed simulations of the motion of the
magnetizations m$_{j}$ of the layers F$_{2}$ of a collection of
different oscillators connected in series. Each $\mathbf{m_{j}}$
is considered as a macrospin without any dipolar interaction with
the other $\mathbf{m_{i}}$. Its time evolution is given by a
Landau-Lifschitz-Gilbert (LLG) equation which includes a standard
spin transfer term proportional to the current. According to
Eq.\ref{hyper}, the current is the sum of the dc current $I_{1}$
plus the coupling term $\sum I_{1} \: \beta_{\Delta Ri} \:
cos(\theta_i(t))$ and the motion equation of $\widehat{m}_{j}$ can
be written as:

\begin{eqnarray}\label{LLG}
\frac{d\widehat{m}_{j}}{dt}
&=&-\gamma_{0}\widehat{m}_{j}\times\overrightarrow{H}_{eff}
+\alpha\widehat{m}_{j}\times\frac{d\widehat{m}_{j}}{dt} \\
&+& \gamma_{0}J[1+\sum_{i=1}^{N} \beta_{\Delta Ri} \:
 cos(\theta_i(t)]\: \widehat{m}_{j}\times(\widehat{m}_{j}\times\widehat{M})\nonumber
\end{eqnarray}

where we have introduced the spin transfer parameter $J$
proportional to $I$ and expressed in field units. In a typical
Co/Cu/Co device\cite{Kiselev}, a current density of 10$^7$
A/cm$^2$ corresponds to about 10$^-$$^2$ Tesla. $\mathbf{M}$ is
the fixed magnetization of all the F$_{1}$ layers. The effective
magnetic field $H_{eff}$, is composed of an uniaxial anisotropy
field $H_{an}$, an applied magnetic field $H_{app}$, and the
demagnetizing field $H_{d}$. All fields are in-plane (parallel to
the direction of the fixed magnetization of F$_{1}$) except for
the out-of-plane demagnetizing field. In the following, if not
mentioned otherwise, we will consider the case of 10 oscillators
with H$_{app}$ = 0.2 T, H$_{d}$ = 1.7 T, and a Gilbert damping
term $\alpha$ = 0.007 (values for Co).

Simulations of the dynamics are performed using a fourth-order
Runge-Kutta algorithm, with a calculation step of 0.5 ps. We have
chosen the following random initial conditions: for each
oscillator, the initial angles between the two magnetizations were
randomly picked between 0 and 10$^{\circ}$ for the polar angle
$\theta_{i}$ and between 0 and 360$^{\circ}$ for the azimuthal
angle $\phi_{i}$. We have checked that, under these conditions,
the variation of the initial conditions does not hinder
synchronization. In order to introduce a dispersion in the
behavior of the oscillators, differences can be introduced in the
anisotropy fields $H_{an}$, demagnetizing fields $H_{d}$ or GMR
ratios. We have checked that all these different types of
dispersion give similar results. In this paper, we will focus on
the first case, with the following dispersion : H$_{an}$ = 0.05 +
(i-1)$\times$ 0.01 in Tesla, \textit{i} varying between 1 and 10.
Finally, in a real experimental setup, the load can be connected
by a few centimeters or meters of microwave cables to the sample
(consisting of closely connected nano-pillars). Delays can occur
in these transmission lines between emission and reception of
microwave currents by the oscillators (differences between the
time \textit{t} when the angles $\theta_i$(t) vary and the time
\textit{t} + $\tau$ when these variations induce current changes).
These delays are taken into account in the present simulations.

We will first consider that all the oscillators have the same
resistance $R$ and magnetoresistance $\Delta R$, so that we can
write the coupling term in Eq.\ref{hyper} as $J A_{GMR}/N \sum
cos(\theta_i(t))$ with A$_{GMR}$ = $\Delta$R/(R + R$_C$/N). In
this particular case, for large N, A$_{GMR}$ is close to the value
of the GMR ratio.

\begin{figure}[h]
   \centering
    \includegraphics[keepaspectratio=1,width=8.5cm]{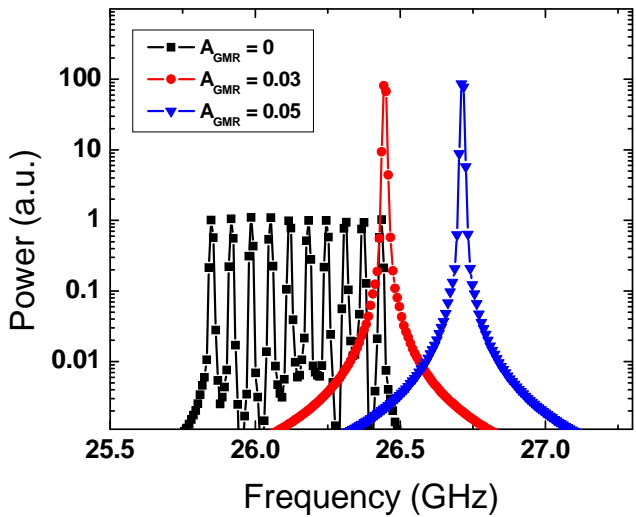}
     \caption{Logarithm of the power versus frequency for the set of
10 oscillators described in the text and for different coupling
     factors A$_{GMR}$. J = 0.035,
and $\tau$ = 5 ps.}
    \label{fig2}
\end{figure}

In Fig.\ref{fig2}, we show the emitted power by the set of 10
oscillators as a function of the frequency for different coupling
parameters A$_{GMR}$. For this set, with 0.05 T $\leq$ H$_{an}$
$\leq$ 0.14 T, the dispersion of the individual frequencies is 2.7
$\%$, of the order of the dispersion (1.25 $\%$) in recent
experiments \cite{Rippard2sync}. The emitted power at a given
frequency is derived by Fast Fourier Transforming the electrical
power released in R$_C$. As the functions $\theta_i(t)$ are known
only for a finite number of times \textit{t}, an oscillation at a
well defined frequency appears in the Fourier transform as a peak
of finite width. The injected dc current J is 0.035 T, and the
delay $\tau$ = 5 ps. When A$_{GMR}$ = 0, each oscillator
oscillates at its own frequency : the frequencies are distributed
between 25.8 and 26.5 GHz approximately, and the total emitted
power is that emitted by the sum of independent oscillators. For
A$_{GMR}$ = 0.03 and 0.05, all the oscillations result in a single
peak. In these two cases, as expected, there is an increase by a
factor of about 100 in the integrated emitted power with respect
to the case without coupling. This scaling with approximately
N$^2$ indicates that the $N$ oscillators are locked not only in
frequency but also in phase, as it will be discussed in more
detail below. In Fig.\ref{fig2}, we can also notice a general
upward shift of the frequency as the coupling increases.

\begin{figure}[h]
   \centering
    \includegraphics[keepaspectratio=1,width=8.5cm]{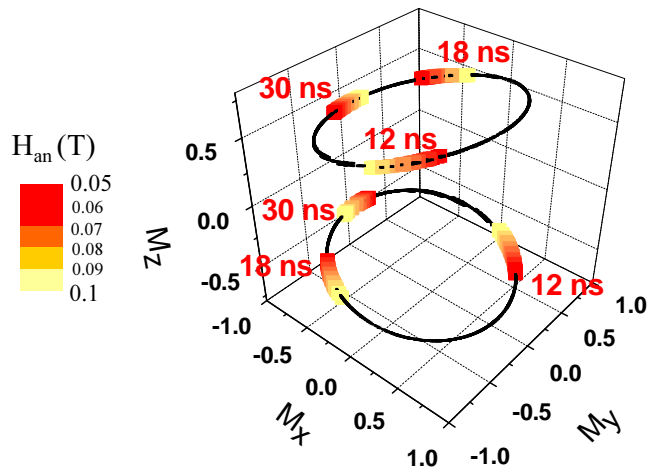}
     \caption{Motion of 100 oscillators on their trajectory at t = 12,
18 and 30 ns : the phase of the oscillators is locked.}
    \label{fig3}
\end{figure}

In Fig.\ref{fig3}, we consider the evolution with time of the
trajectories of 100 oscillators. The bias conditions (J = 0.035 T,
$\tau$ = 5 ps) are similar to the previous case with H$_{an}$
picked randomly between 0.05 and 0.1 T for each oscillator (see
scale on Fig.\ref{fig3}), and A$_{GMR}$ = 0.03. The black curve
corresponds to the two symmetrical final trajectories. By looking
at the position of the oscillators at different times, we see they
are turning in phase (small bounded phase shift), with the fastest
oscillator opening the way.

\begin{figure}[h]
   \centering
    \includegraphics[keepaspectratio=1,width=8.5cm]{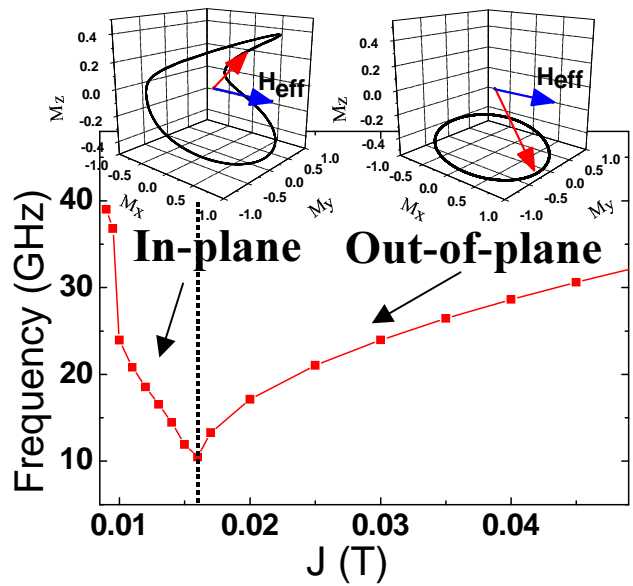}
     \caption{Frequency versus injected dc current J for
       oscillator 1 and A$_{GMR}$ = 0 (no coupling). The insets show the
       trajectories of the magnetization $\widehat{m}$ (M$_{x}$,M$_{y}$,M$_{z}$)
       in the two regimes referred to in the text. In the absence of current, the equilibrium
       along $H_{eff}$ corresponds to M$_{x}$ = 1, M$_{y}$ = M$_{z}$ =
       0.}
    \label{fig4}
\end{figure}

\begin{figure}[h]
   \centering
    \includegraphics[keepaspectratio=1,width=8.5cm]{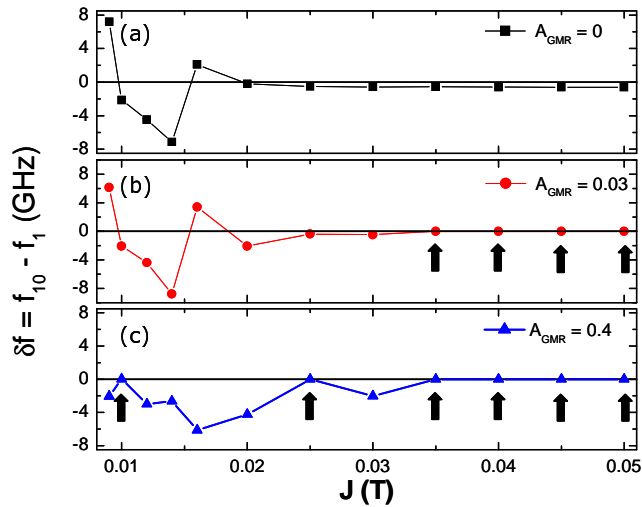}
     \caption{Difference in frequency between oscillator 1 (H$_{an}$ = 0.05 T)
     and oscillator 10 (H$_{an}$ = 0.14 T) as a function of the dc current. Three cases are
      considered. (a) A$_{GMR}$ = 0, (b) A$_{GMR}$ = 0.03 and $\tau$ = 5 ps,
       (c) A$_{GMR}$ = 0.4 and $\tau$ = 0.3 ns. The black
       arrows indicate synchronization (f$_{10}$ - f$_1$ = 0).}
    \label{fig5}
\end{figure}

Experimentally, varying the coupling parameter $A_{GMR}$ means
changing the GMR ratio in a controllable way, which might be
difficult. Another way to increase the coupling is to increase J
 which, from Eq.\ref{LLG}, enhances both the mean torque and the coupling
  between \textit{i} and \textit{j}.
  The variation of the frequency of
oscillator 1 (H$_{an}$ = 0.05 T) with J in the absence of coupling
($A_{GMR}$ = 0) is shown on Fig.\ref{fig4}. Similar results have
been obtained in simulations by other groups\cite{Kiselev}. For J
smaller than 0.016 T, the frequency decreases as J increases in
the regime of in-plane precessional trajectories of the
magnetization. It increases for J larger than 0.016 T
corresponding to the regime of out-of-plane orbits. In
Fig.\ref{fig5}, we have plotted the difference in frequency,
$\delta$f, between the 10$^{th}$ oscillator and the 1$^{st}$ as a
function of J for different coupling parameters A$_{GMR}$;
$\delta$f = 0 means synchronization of the two oscillators. The
reference curve (no synchronization) obtained for A$_{GMR}$ = 0 is
plotted in Fig.\ref{fig5}(a).

We first consider the curve of Fig.\ref{fig5}(b) corresponding to
A$_{GMR}$ = 0.03 with a delay $\tau$ of 5 ps. For low values of
$J$, the coupling is small, and the oscillators do not
synchronize. The system is nevertheless disturbed by the injection
of the microwave currents, as can be seen from the differences
between $\delta$f for A$_{GMR}$ = 0.03 and A$_{GMR}$ = 0.
Synchronization is reached ($\delta$f = 0, see arrows on
Fig.\ref{fig5}(b)) above J = 0.035 T  (this is in the out-of-plane
regime with, as shown on Fig.\ref{fig5}(a), weaker dispersion, and
probably, easier synchronization). Fig.\ref{fig5}(c) corresponds
to a situation with enhanced coupling (larger A$_{GMR}$). In this
case, synchronization extends to the in-plane precession regime
(see arrow at J = 0.01 T).

Another important parameter acting on the synchronization is the
delay time $\tau$. We have studied the number of synchronized
oscillators out of ten as a function of the delay $\tau$, with J =
0.035 T, A$_{GMR}$ = 0.1 and H$_{an}$ = 0.01 + (i-1) $\times$ 0.01
T. For values of $\tau$ smaller than about 0.25 ns, the number of
synchronized oscillators varies periodically with $\tau$, with a
period of approximately 39 ps. As the frequency at synchronization
is about 27.5 GHz which corresponds to a period $T$ of 37 ps,
there is a link between the period $T$ of the oscillators and the
period of the oscillatory variation as a function of $\tau$. The
oscillators can be synchronized more easily when they receive at
time \textit{t} microwave currents emitted at time \textit{t-nT}
(n = integer), since at \textit{t-nT} they were in the same state
as at \textit{t}. \cite{Strogatzdelay}. This periodic behavior of
synchronization versus delay, with a period corresponding to the
oscillation frequency, has been already predicted in other systems
ruled by different sets of equations
\cite{Rosenblum,Strogatzdelay}. The best conditions are for $\tau$
between 0.2 and 0.75 ns and the proportion of synchronized
oscillators decreases again above 0.8 ns.

In conclusion, we have shown that it is possible to synchronize a
network of spin-transfer oscillators by simply connecting them
electrically in series to a load (similar effects can be expected
for oscillators in parallel). The synchronization depends on the
dispersion of the individual frequencies, on the coupling
parameters and the delay time $\tau$. Under certain conditions,
the synchronization can be complete. In this case, the output
power of N oscillators turns out to scale with N$^2$. We have also
shown that, for synchronized oscillators, the frequency as well as
the emitted power are strongly dependent on the coupling factor
A$_{GMR}$, related to the GMR ratio. These results are of interest
for obtaining an enhanced microwave generation with networks of
spin-transfer oscillators. They also show that magnetic devices
can be synchronized in the same way (from the coupling mechanism
point of view) as in the model system represented by a network of
Josephson junctions, but with two degrees of freedom (polar and
azimuthal angles) instead of one (phase).

We finally point out that the synchronization mechanism by
microwave current components we have discussed for networks could
also be important in the interpretation of the properties of a
single spin-transfer oscillator (pillars or point contacts). The
microwave spectrum of some isolated oscillators is surprisingly
narrow, in contrast with the inhomogeneous broadening predicted by
simulations based on micromagnetic models of ferromagnetic
dots\cite{micromag}. However, from our results, introducing the
coupling between different parts of the dot due to the microwave
component of the total current could synchronize these different
parts. Such synchronization effects could thus explain less
chaotic oscillations than predicted and account for the narrow
linewidth of the microwave spectra.


\vspace{.5cm}


\begin{thebibliography}{31}

\bibitem{Slonc}
J. Slonczewski, J. Magn. Magn. Mater. {\bf 159}, L1 (1996)

\bibitem{Tsoi}
M. Tsoi \textit{et al}., Phys. Rev. Lett. {\bf 80}, 4281 (1998)

\bibitem{Wegrove}
J. E. Wegrove \textit{et al}., Europhys. Lett. 45, 626 (1999)

\bibitem{Katine}
J. A. Katine \textit{et al.}, Phys. Rev. Lett. {\bf 84}, 3149,
(2000)

\bibitem{JulieAPL}
J. Grollier \textit{et al}., Appl. Phys. Lett. {\bf 78}, 3663
(2001)

\bibitem{Myers}
E. B. Myers \textit{et al}., Phys. Rev. Lett. {\bf 89}, 196801
(2002)

\bibitem{Ozyilmaz}
B. Ozyilmaz \textit{et al}., Phys. Rev. Lett. {\bf 91}, 067203
(2003)

\bibitem{Urazdhin}
S. Urazhdin, N. O. Birge, W. P. Pratt, Jr., and J. Bass, Phys.
Rev. Lett. {\bf 91}, 146803, (2003)

\bibitem{KiselevPRL}
S. I. Kiselev \textit{et al}., Phys. Rev. Lett. {\bf 93}, 036601
(2004)

\bibitem{Berger}
L. Berger, J. Magn. Magn. Mater. {\bf 278}, 185 (2004)


\bibitem{Stiles}
M. D. Stiles, Jiang Xiao, A. Zangwill, Phys. Rev. B {\bf 69},
054408 (2004)

\bibitem{Zhang}
J. Zhang, P. M. Levy, S. Zhang, V. Antropov, Phys. Rev. Lett. {\bf
93}, 256602 (2004)

\bibitem{Manschot}
J. Manschot, A. Brataas, G. E. W. Bauer, Appl. Phys. Lett. {\bf
85}, 3250 (2004)

\bibitem{Fert}
A. Fert \textit{et al}., J. Magn. Magn. Mater. {\bf 272}, 1706
(2004)

\bibitem{Barnas}
J. Barnas \textit{et al.}, cond-mat/0501570 (2005)

\bibitem{Thibault}
T. Devolder \textit{et al.}, Appl. Phys. Lett. {\bf 86}, 062505
(2005)

\bibitem{Kiselev}
S. I. Kiselev, J.C. Sankey, I.N. Krivorotov, N.C. Emley, Nature
{\bf 425}, 380 (2003)

\bibitem{Rippard}
W. H. Rippard et \textit{al.}, Phys. Rev. Lett. {\bf 92}, 027201
(2004).


\bibitem{Krivotorov}
I. N. Krivorotov \textit{et al.}, Science {\bf 307}, 228 (2005)

\bibitem{Baibich}
M. N. Baibich \textit{et al}., Phys. Rev. Lett. {\bf 61}, 2472
(1988).


\bibitem{RippardPRB}
W. H. Rippard \textit{et al.}, Phys. Rev. B {\bf 70}, 100406(R)
(2004)



\bibitem{StrogatzNonLin}
S. H. Strogatz, Nonlinear Dynamics and Chaos, Addison-Wesley
Publishing Company (1994)

\bibitem{Winfree}
A. T. Winfree, The Geometry of Biological Time (Springer, New
York, 1980)

\bibitem{Newrock}
R. S. Newrock, C. J. Lobb, U. Geigenmuller, M. Octavio, Solid
State Phys. {\bf 54}, 263 (2000)

\bibitem{StrogatzJosephson}
K. Wiesenfeld, P. Colet, S. H. Strogatz, Phys. Rev. E {\bf 57},
1563 (1998)

\bibitem{JosephExp}
P. Barbara, A. B. Cawthorne, S. V. Shitov, C. J. Lobb, Phys. Rev.
Lett. {\bf 82}, 1963 (1999); B. Vasilic, S. V. Shitov, C. J. Lobb,
P. Barbara, Appl. Phys. Lett. {\bf 78}, 1137 (2001)

\bibitem{Duzer}
T. V. Duzer, C. W. Turner, Superconductive Devices and Circuits
(Prentice Hall, Upper Saddle River, NJ) (1999)

\bibitem{RippardLock}
W.H. Rippard \textit{et al.}, Phys. Rev. Lett. {\bf 95}, 067203
(2005)

\bibitem{Slavin}
A.N. Slavin, V.S. Tiberkevich, to be published in Phys. Rev. B

\bibitem{Rippard2sync}
S. Kaka \textit{et al.}, Nature {\bf 437}, 389 (2005)

\bibitem{Mancoff2sync}
F.B. Mancoff, N.D. Rizzo, B.N. Engel, S. Tehrani, Nature {\bf
437}, 393 (2005)

\bibitem{Strogatzdelay}
M. K. S. Yeung, S. H. Strogatz, Phys. Rev. Lett. {\bf 82} , 648
(1999)

\bibitem{Rosenblum}
M. G. Rosenblum, A. S. Pikovsky, Phys. Rev. Lett. {\bf 92}, 114102
(2004)

\bibitem{micromag}
K. J. Lee \textit{et al.}, Nat. Mat. {\bf 3}, 877 (2004); X. Zhu,
J.-G. Zhu, R. M. White, J. Appl. Phys. {\bf 95}, 6630 (2004); B.
Montigny, J. Miltat, J. Appl. Phys. {\bf 97}, 10C708 (2005); D.V.
Berkov, N.L. Gorn, cond-mat/0503754 (2005)

\end{thebibliography}
\end{document}